\documentclass[referee]{aa}
\usepackage{graphicx}

\begin{document}
\title{Distances to Nearby Galaxies in Sculptor
\thanks{Based on observations made with the NASA/ESA Hubble Space
Telescope.  The Space Telescope Science Institute is operated by the
Association of Universities for Research in Astronomy, Inc. under NASA
contract NAS 5--26555.}}
\titlerunning{Sculptor group of galaxies }
\author{I. D. Karachentsev \inst{1}
\and E. K. Grebel \inst{2}
\and M. E. Sharina \inst{1,10}
\and A. E. Dolphin \inst{3}
\and D.~Geisler \inst{4}
\and P. Guhathakurta \inst{5}
\and P.~W. Hodge \inst{6}
\and V. E. Karachentseva \inst{7}
\and A.~Sarajedini \inst{8}
\and P. Seitzer \inst{9}}
\institute{Special Astrophysical Observatory, Russian Academy
of Sciences, N. Arkhyz, KChR, 369167, Russia
\and Max-Planck-Institut f\"{u}r Astronomie, K\"{o}nigstuhl 17, D-69117 
Heidelberg, Germany
\and Kitt Peak National Observatory, National Optical Astronomy
Observatories, P.O. Box 26732, Tucson, AZ 85726, USA
\and Departamento de F\'{\i}sica, Grupo de Astronom\'{\i}a, Universidad de
Concepci\'on, Casilla 160-C, Concepci\'on, Chile
\and UCO/Lick Observatory, University of California at Santa Cruz, Santa
Cruz, CA 95064, USA
\and Department of Astronomy, University of Washington, Box 351580,
Seattle, WA 98195, USA
\and Astronomical Observatory of Kiev University, 04053, Observatorna 3,
Kiev, Ukraine
\and Department of Astronomy, University of Florida, Gainesville, FL
32611, USA
\and Department of Astronomy, University of Michigan, 830 Dennison
Building, Ann Arbor, MI 48109, USA
\and Isaac Newton Institute, Chile, SAO Branch}

\date{Received:  December 10, 2002}
%\maketitle

\abstract{
We present an analysis of Hubble Space Telescope/WFPC2 images of nine
nearby galaxies in Sculptor. We derive their distances from the luminosity
of the tip of the red giant branch stars with a typical accuracy of
$ \sim 12$ \%. Their distances are 4.21 Mpc (Sc~22), 4.92 Mpc (DDO~226),
3.94 Mpc (NGC~253), 3.40 Mpc (KDG~2), 3.34 Mpc (DDO~6), 3.42 Mpc (ESO~540-030),
4.43 Mpc (ESO~245--05), 4.27 Mpc (UGCA~442), and 3.91 Mpc (NGC~7793). The
galaxies are concentrated in several spatially separated loose groups around
NGC~300, NGC~253, and NGC~7793. The Sculptor galaxy complex together with
the CVn I cloud and the Local Group form a 10 Mpc amorphous filament, apparently
driven by the free Hubble flow.
}
\maketitle

\section{Introduction}

 The association of five bright spiral galaxies in the Sculptor constellation,
NGC~55, NGC~247, NGC~253, NGC~300, and NGC~7793, has been considered by many
authors as the galaxy group closest to the Local Group (LG), at a distance of
$\sim$2 Mpc. Over the past decades, a number of dwarf galaxies  have
been found in this area by van den Bergh (1959: DDO~6, DDO~226), Nilsen
(1974: UGCA~438, UGCA~442), and Lauberts (1982: ESO sky survey objects).
Special searches for fainter irregular (dIrr) and spheroidal (dSph)
dwarf galaxies in the Sculptor group were undertaken by C\^{o}t\'e et al. 
(1997), Karachentseva \& Karachentsev (1998, 2000), and Jerjen et al. (1998, 
2000).  In their paper entitled,
``Discovery of numerous dwarf galaxies in the two nearest groups of
galaxies'', C\^{o}t\'e et al. (1997) reported on the discovery of five new
dwarf members of the Sculptor group: Sc~2, Sc~18, Sc~22, Sc~24, and Sc~42.
 However, based on CCD images of the objects obtained by Whiting et al.(2002)
and Tully (2003), we conclude that all of them, except Sc~22,
are background galaxies, whose
H\,{\sc i} velocities were confused with Galactic high velocity clouds
in the velocity range from +60 to +160 km s$^{-1}$. Recent H\,{\sc i} surveys 
of the Sculptor region by Staveley-Smith et al. (1998), Barnes et al. (2001),
de Blok et al. (2002) have not yet led to the discovery of new dwarf members
of the group. Jerjen et al. (1998) measured distances to five dwarf galaxies
in Sculptor from fluctuations of their surface brightness and showed that 
the so-called ``Sculptor group'' turns out to be a loose filament of galaxies
extended along a line of sight over $\sim$5 Mpc.

  In this paper we present new distance measurements for nine galaxies in
the Sculptor region, derived from their tip of the red giant branch (TRGB).
Together with earlier published TRGB distances to four other galaxies
(Karachentsev et al. 2000, 2002) and the distance to NGC~300 from Cepheids
(Freedman et al. 1992), this gives us a basis for a more detailed study of
the structure of the Sculptor complex.

\section { WFPC2 photometry}

   Images of nine galaxies were obtained with the Wide Field and Planetary
Camera (WFPC2) aboard the Hubble Space Telescope (HST) between August 24,
1999 and June 28, 2001 as  part of our HST snapshot survey of nearby galaxy
candidates (Seitzer et al. 1999; Grebel et al. 2000). The galaxies were
observed with 600-second exposures taken in the F606W and F814W filters
for each object.  Digital Sky Survey (DSS) images of these galaxies are 
shown in Figure 1 with the HST WFPC2 footprints superimposed. The field size
of the red DSS-II images is 8$\arcmin$. Small galaxies were usually
centered on the WF3 chip, but for some bright objects the WFPC2
position was shifted towards the galaxy periphery to decrease
stellar crowding. The WFPC2 images of the galaxies are presented
in the upper panels of Figure 2, where both filters are combined.

 For photometric measurements we used the HSTphot stellar photometry
package developed by Dolphin (2000a). The package has been optimized
for the undersampled conditions present in the WFPC2 to work in
crowded fields.  After removing cosmic rays,
simultaneous photometry was performed on the F606W and F814W frames
using \textit{multiphot}, with corrections to an aperture with a
radius of $0\farcs5$. Charge-transfer efficiency (CTE) corrections and
calibrations were then applied, which are based on the Dolphin (2000b)
formulae, producing $V, I$ photometry for all stars detected in both
images. Additionally, stars with a signal-to-noise ratio $S/N < 3 $,
$\mid \chi \mid\,\, >2.0$, or  $\mid$ sharpness $\mid\,\, >0.4$  in each
exposure were eliminated from the final photometry list. The uncertainty
of the photometric zero point is estimated to be within $0\fm05$
(Dolphin 2000b).

\section{TRGB distances to nine galaxies in Sculptor}

The tip of red giant branch (TRGB) method provides an efficient tool
to measure galaxy distances. The TRGB distances agree with those given by
the Cepheid period-luminosity relation to within 5\%. As shown by
Lee et al. (1993), the TRGB is relatively independent
of age and metallicity. In the $I$ band the TRGB for low-mass stars
is found to be stable within $ \sim 0\fm1$  (Salaris \& Cassisi 1997;
Udalski et al. 2001) for metallicities, [Fe/H], encompassing the entire
range from $-$2.1 to $-$0.7 dex found in Galactic globular clusters.
According to Da Costa \& Armandroff (1990), for metal-poor systems the TRGB
is located at $M_I = -4\fm05$. Ferrarese et al. (2000) calibrated the
zero point of the TRGB from galaxies with Cepheid distances and estimated
$M_I = -4\fm06 \pm0\fm07(random)\pm0.13(systematic)$. A new TRGB calibration,
$M_I = -4\fm04 \pm0\fm12$, was made by Bellazzini et al. (2001) based on
photometry and on the distance estimate from a detached eclipsing binary in the
Galactic globular cluster $ \omega$ Centauri. For this paper we use
 $M_I = -4\fm05$. The lower left panels of Figure 2 show
$I$, $(V-I)$ color-magnitude diagrams (CMDs) for the nine
observed galaxies.

   We determined the TRGB using a Gaussian-smoothed $I$-band luminosity
function (LF) for red stars with colors $V-I$ within $\pm0\fm5$ of the mean
$\langle V-I \rangle$ expected for red giant branch stars. Following
Sakai et al. (1996), we applied a Sobel edge-detection filter.
The position of the TRGB was identified with the peak in the
filter response function. The resulting LFs and the Sobel-filtered LFs
are shown in the lower right corners of Figure 2. The results are
summarized in Table 1. There we list:
(1) galaxy name; (2) equatorial coordinates of the galaxy center;
(3) galaxy major diameter and axial ratio; (4) apparent integrated blue
magnitude from the NASA Extragalactic Database (NED) and Galactic
extinction in the  $B$-band from Schlegel et al. 1998; (5) morphological
type in de Vaucouleurs' notation; (6) heliocentric radial velocity and
radial velocity with respect to the LG centroid
(Karachentsev \& Makarov 1996); (7) position of the TRGB and its
uncertainty as derived with the Sobel filter; (9) true distance modulus
with its uncertainty, which takes into account the uncertainty in the TRGB,
as well as uncertainties of the HST photometry zero point ($\sim0\fm05$),
the aperture corrections ($\sim0\fm05$), and the crowding effects
($\sim0\fm06$) added quadratically; the uncertainties in the extinction and
reddening are taken to be $10\%$ of their values from Schlegel et al. (1998);
[for more details on the total budget of internal and external systematic
errors for the TRGB method see Mendez et al. (2002)]; and (9) linear distance
in Mpc and its uncertainty.

   Given the distance moduli of the galaxies, we can estimate their
mean metallicity  from the mean color of the TRGB measured at an
absolute magnitude $ M_I = -3.5$, as recommended by Da Costa \& Armandroff
(1990). Based on a Gaussian fit to the color distribution of the giant
stars in a corresponding $I$-- magnitude interval $ (-3.5\pm 0.3)$, we derived
their mean colors, $ (V - I)_{-3.5} $, which lie in the range of 1.18 mag to
1.54 mag
after correction for Galactic reddening. Following the relation of Lee et al.
(1993), this provides us with mean metallicities 
$-$1.1 dex $> \langle$[Fe/H]$\rangle > -2.4$ dex, 
listed in the last column of Table 1. With a typical statistical scatter
of the mean color ($\sim0\fm05$), and uncertainties of the HST photometry
zero point we expect the uncertainty in metallicity to be about 0.3 dex.
Therefore within the measurement accuracy the metallicity of the galaxies
satisfy the required limitation, [Fe/H] $< -0.7$ dex. Below, some
individual properties of the galaxies are briefly discussed.

  {\em Sc 22.} This dwarf spheroidal galaxy of very low surface brightness
was discovered by C\^{o}t\'e et al. (1997). Surface photometry of Sc~22 
was carried
out by Jerjen et al. (1998, 2000), who determined its integrated apparent
magnitude, $B_T$ = 17.73 mag, integrated color, $(B-R)_T$ = 0.79 mag, and 
central
surface brightness, $25.8^m/\sq\arcsec$ in the B band. Using the method of 
surface
brightness fluctuations (SBF), they estimated the distance to Sc~22 to be
2.67$\pm$0.16 Mpc. The galaxy was not detected in the H\,{\sc i} line by
C\^{o}t\'e et al. (1997), neither was it detected in the ``blind'' HIPASS survey
by Staveley-Smith et al. (1998) and Barnes et al. (2001). The color-magnitude
diagram of Sc 22 (see Fig. 2) is populated predominantly by red stars.
The application of the Sobel filter yields I(TRGB) = $24\fm10\pm0\fm21$,
which corresponds to a linear distance of 4.21$\pm$0.43 Mpc.  This is
much larger than the SBF distance.

  {\em DDO 226 = IC 1574 = UGCA 9.} This dIrr galaxy has a relatively
high radial velocity, $V_{LG}$ = 408 km s$^{-1}$. The CMD of DDO~226 shows
blue and red stellar populations.  There is no strong discontinuity in
the luminosity function but there is only a slight hint of a red giant branch.
Two peaks are seen in the Sobel--filtered luminosity function. The first lower
peak at $I = 23.8$ appears to be caused by AGB stars, and the second one, at
$I = 24\fm44\pm0\fm24$, which we interpret as the TRGB, yields a linear
distance of 4.92$\pm$0.58 Mpc.

 {\em NGC 253.} NGC 253 is the brightest galaxy in the Sculptor group.
With a size of 27$\arcmin\times6\arcmin$, NGC~253 extends far beyond
the WFPC2 field. Surprisingly, this prominent Sc galaxy has so far no
reliable distance estimate, apart from a rough estimate of $D$ = 2.77 Mpc
derived by Puche \& Carignan (1988) from the Tully-Fisher relation.
In our HST observations the WFPC2 was pointed to the NE  quadrant of the
galaxy, where the stellar crowding is lower. The HST photometry measured
a total of about 27000, mostly red, stars. The TRGB is located at
$I$ = 23$\fm97\pm0\fm19$, corresponding to a distance of 3.94$\pm$0.37 Mpc.
With this distance, the resulting absolute integrated magnitude of NGC 253,
$M_B = -20.14$ mag, turns out to be comparable with the absolute magnitudes of
the Milky Way.Like the Milky Way, NGC 253 has a rotation velocity
$V_{max}$ about 225 km s$^{-1}$.

  {\em KDG 2 = ESO 540--030 = KK 09.} This dSph galaxy of low surface
brightness was found by Karachentseva (1968) and then selected as
the Local Volume member candidate by Karachentseva \& Karachentsev (1998).
According to the surface photometry carried out by Jerjen et al. (1998,
2000), KDG 2 has an integrated magnitude of $B_T$ = 16.37 mag, an integrated
color of $(B-R)_T$ = 0.83 mag, and a central surface brightness of
24.1$^m/\sq\arcsec$ in the $B$-band. Jerjen et al. (1998) determined its
distance via SBF to be 3.19$\pm$0.13 Mpc. The HST photometry measured
about 1900 predominantly red stars. We derived the TRGB to be located at
$23\fm65\pm0\fm20$, which corresponds to $D$ = 3.40$\pm$0.34 Mpc, in good
agreement with the previous estimate. Apart from red stars, we also found
a number of blue stars, which occupy the central part of the galaxy.
Probably, KDG~2 is not a dSph galaxy, but belongs to a transition
dSph/dIrr type, like LGS-3 and Antlia. Huchtmeier et al. (2000) did not
detect it in the H\,{\sc i} line.

  {\em DDO 6 = UGCA 15 = ESO 540--031.} This dIrr galaxy of drop-like shape
has a radial velocity of $V_{LG}$ = 348 km s$^{-1}$. In Fig.2 the CMD
shows the presence of mixed blue and red populations, in particular,
a prominent population 
of RGB stars. We determined I(TRGB) to be $23\fm62\pm0\fm16$, yielding
$D$ = 3.34$\pm$0.24 Mpc.

  {\em ESO 540-032 = FG 24 = KK 10.} Like KDG 2, FG 24 (Feitzinger \& Galinski
1985) has a reddish color and a low surface brightness, typical of dSph
or dSph/dIrr galaxies. The galaxy was not detected in H\,{\sc i} by Huchtmeier 
et al. (2000). Surface photometry of FG~24, performed by Jerjen et al. (1998,
2000), yields an integrated magnitude of $B_T$ = 16.44 mag, an integrated color
$(B-R)_T$ = 1.08 mag, and a central surface brightness 24.5$^m/\sq\arcsec$ 
in the $B$-band. Most of the stars detected by us in FG~24 (see Fig. 2) are 
likely RGB stars, although some faint blue stars are present in the galaxy's 
central
part. The TRGB $I$-band magnitude derived by us, $23\fm67\pm0\fm17$, 
corresponds to a distance
of 3.42$\pm$0.27 Mpc, which exceeds the distance 2.21$\pm$0.14 Mpc
estimated by Jerjen et al. (1998) from SBF, but agrees excellently with
the new distance estimate, 3.4$\pm$0.2 Mpc, obtained by Jerjen \& Rejkuba 
(2001) from the TRGB.

  {\em ESO 245--005.} This irregular galaxy of Magellanic type has an angular
dimension of $3\farcm8\times3\farcm4$  and a radial velocity of 
$V_{LG}$ = 308 km s$^{-1}$. In
the WFPC2 image of its central bar-like part we detected about
11000 blue and red stars. We determined the $I$-band TRGB to be at
24$\fm22\pm0\fm21$, yielding a distance of 4.43$\pm$0.45 Mpc.

  {\em UGCA 442 = ESO 471--06.} This is an edge-on galaxy of Im type with a
radial velocity $V_{LG}$ = 299 km s$^{-1}$ and angular dimension 
6$\farcm4\times0\farcm9$,
much larger than the WFPC2 field. The galaxy seems to be well resolved
into stars. Our HST photometry reveals about 5600 stars, both blue and red
ones. The derived TRGB, $I = 24\fm13\pm0\fm26$, corresponds to a distance
of 4.27$\pm$0.52 Mpc.

  {\em NGC 7793.} This is a bright spiral galaxy with a dimension of
9$\farcm3\times6\farcm3$ and with a radial velocity $V_{LG}$ = 252 km s$^{-1}$. The WFPC2 was pointed
at its eastern side. We obtained photometry for about 22000 stars. 
The CMD shows a mixture of young and old stellar populations.
The TRGB magnitude, 23$\fm95\pm0\fm22$, yields a galaxy distance of
3.91$\pm$0.41 Mpc, in reasonable agreement with the distance 3.27$\pm$0.08 Mpc
derived by Puche \& Carignan (1988) from the Tully-Fisher relation.

  \section {Structure and kinematics of the Sculptor group}

  To study the 3-D structure of the Sculptor complex, we collected the
most complete sample of data on all known nearby galaxies situated
in this region of the sky. Table 2 presents the following characteristics of the 21
galaxies in our sample: (1) galaxy name; (2) equatorial (upper line), and
Supergalactic (lower line) coordinates; (3) major angular diameter in
arcmin and apparent axial ratio; (4) apparent integrated magnitude from
NED and Galactic extinction from Schlegel et al. (1998) in the B-band;
(5) morphological type; (6) heliocentric radial velocity in km s$^{-1}$ and
velocity in the Local Group rest frame (Karachentsev \& Makarov 1996);
(7) the H\,{\sc i} line width (in km s$^{-1}$) at the 50\% level of the 
maximum from LEDA
(Paturel et al. 1996) or HIPASS, corrected for galaxy inclination (upper
line), and absolute magnitude of the galaxy, corrected for Galactic
extinction (lower line); (8) distance to the galaxy in Mpc from the
Milky Way (upper line), and from the Local Group centroid (lower line),
respectively. The last column gives the method used for distance
measurement (``Cep'' --- from Cepheids, ``RGB'' --- from TRGB, ``SBF'' --- from
surface brightness fluctuations, and ``TF'' --- from the Tully-Fisher relation),
and the distance data reference.

  Apart from four dSph galaxies with known distances but unknown radial
velocities (KK~3, Sc~22, KDG~2, and FG~24), we also included in Table 2 
the low surface brightness galaxy KK~258,
 which looks like a nearby semi-resolved system in a CCD image
obtained by Whiting et al. (2002).
Its distance and velocity are both unknown.
In comparison with the TRGB distances, the distance estimates of
Sc~22, KDG~2 and FG~24 from SBF (Jerjen et al. 1998) are systematically
lower on 0.9 Mpc, which is why we increased their original SBF distance
estimate for NGC~59 from 4.4 Mpc to 5.3 Mpc.

 The distribution of the 21 galaxies from Table 2 is presented in Fig. 3 in
equatorial coordinates. Three of the brightest spiral galaxies are shown
as filled squares, and dIrr and dSph galaxies are shown as filled and
open circles, respectively. Radial velocities of the galaxies with respect
to the LG centroid are indicated by numbers. As can be seen, the distribution
of galaxies in Sculptor does not exhibit any distinct center, and
it does not show a sharp boundary either.
Most of the galaxies are situated at low supergalactic latitudes,
$\mid$SGB$\mid$ $< 10\degr$, where projection effects makes it difficult
to restore the 3-D structure of the group. Referring to Table~3, we
recognize that the spiral galaxy NGC 253 surpasses all
other galaxies in this group by more than a factor of five in 
luminosity. Therefore, we consider NGC~253 as the
dynamical center of the Sculptor group.

  Puche \& Carignan (1988) estimated the distance to NGC~253 and some other
galaxies in Sculptor, based on the Tully-Fisher relation. However, for
their calibration, they used old data on galaxy distances. A revised relation
``absolute magnitude --- H\,{\sc i} line width'' for 6 galaxies in Sculptor 
with new distance estimates is presented in Fig.~4. Here, the line width is
corrected for galaxy inclination to the line of sight. Using the regression
line, $M_B = -7.0$ log$(W_c) - 1.8$, we estimated the distances to five 
galaxies:
NGC~55, NGC~247, NGC~625, ESO~349--031, and ESO~149--03, presented in
column (8) of Table 2.

  The Hubble diagram showing velocity versus distance for 20 galaxies 
in Sculptor
is given in Fig.~5.  Galaxies with accurate distance estimates (``Cep'',
``RGB"'', ``SBF'') are shown as filled circles, while galaxies with distances
from the T-F relation are indicated by crosses, and the brightest spiral
galaxy NGC~253 is shown as a square. Four dSph galaxies without velocities
are shown conditionally by vertical bars. The solid  curve corresponds
to the Hubble parameter H = 75 km s$^{-1}$ Mpc$^{-1}$. At small distances this curve
deviates from a straight line due to a decelerating gravitational action
of the Local Group, the total mass of which is adopted to be
$M_{LG} = 1.3\cdot 10^{12}\cdot M_{\sun}$ (Karachentsev et al. 2002).

  Based on the data in Fig.\ 3 and Fig.\ 5, we can describe the structure of
the galaxy complex in Sculptor in the following way.

  a) Three galaxies with $V_{LG} >$ 400 km s$^{-1}$ (ESO~149--03, NGC~59, and DDO~226)
apparently are background objects.  NGC~59 and DDO~226 probably form
a wide pair with a linear projected distance of 580 kpc and a radial
velocity difference of 13 km~s$^{-1}$.

  b) In front of the complex there is a pair of bright galaxies, NGC~300
and NGC~55, with two dSph companions, ESO~410--05 and ESO~294--10. The
mean distance to this loose quartet is 1.95 Mpc.  This is only 0.6 Mpc
more distant than to another known loose quartet: NGC~3109, Sex~A, Sex~B, and
Antlia, located at the LG edge. Possibly, two other dIrr galaxies, UGCA~438
and IC~5152, are associated with the NGC~300 group, too.

  c) The brightest spiral galaxy NGC~253 together with its companions,
NGC~247, DDO~6, Sc~22, KDG~2 and FG~24, can be considered as the Sculptor
complex' core. The last three dSph galaxies do not have radial velocities so
far. Apart from the NGC~253 group, there is a galaxy triplet, NGC~7793,
UGCA~442, and ESO~349--031. The line of sight distances to
NGC~253 and NGC~7793 are almost the same within the measurement errors.

  d) Among the remaining galaxies, NGC~625 and ESO~245--05 have an angular
separation of 2$\fdg9$ and a radial velocity difference of only 27 km s$^{-1}$.
However, their distance estimates differ significantly. Because
NGC~625 deviates essentially from the Hubble regression line in Fig.\ 5,
we assume its T-F distance to be underestimated by approximately 2 Mpc.
This assumption can be easily proven by measuring the TRGB distance
to this poorly studied galaxy.

  The total mass of each group can be estimated from the virial balance
of kinetic and potential energies (Limber \& Mathews 1960),
$$M_{vir} = 3\pi N \cdot(N-1)^{-1} \cdot G^{-1} \cdot \sigma^2_v \cdot R_H,
\eqno(1)$$
where $N$ is the number of galaxies in the group, $\sigma_v^2$ is the 
radial velocity dispersion,  $R_H$ is the mean projected harmonic radius, and
$G$ is the gravitational constant. Such an approach assumes that the
characteristic crossing time of the group, $T_{cross} = \langle R_p\rangle 
/\sigma_v$,
is low in comparison with the age of the Universe (here $\langle R_p\rangle$ 
means the average projected radius of the group).

  Another way to estimate the total mass of a group was proposed by
Bahcall \& Tremaine (1981). Assuming the motions of dwarf galaxies around the
main group member to be closed Keplerian motions with orbit eccentricity
$e$, in the case of random orientation of galaxy orbits we obtain
$$M_{orb} = (32/3\pi)\cdot G^{-1}\cdot (1- 2e^2/3)^{-1} \langle R_p\cdot 
\Delta V^2_r \rangle, \eqno(2) $$
where $R_p$ and $\Delta V$ are  projected distance and radial velocity of
a companion with respect to the main group member.

  The basic dynamical parameters of the three mentioned groups in Sculptor
are presented in Table 3. In the case of orbital mass estimates the mean
eccentricity  $e$= 0.7 is adopted. As seen from these data, the
virial/orbital mass-to-luminosity ratios of the groups lie in the range
of 45 to 260 $M_{\sun}/L_{\sun}$. However, for all the groups their crossing
time, 6 -- 18 Gyr, is comparable to the time of cosmic expansion, $1/H_0$,
which makes the derived mass estimates very unreliable.

  According to Lynden-Bell (1981) and Sandage (1986), in the expanding
universe any dense enough group with a total mass  $M_0$ may be
characterized by a ``zero-velocity surface'', which separates the group
from the Hubble flow. In the case of spherical symmetry, the radius of
this surface, $R_0$, is expressed via the total mass of the group and
the Hubble constant, $H_0$, by a simple relation
      $$M_0 = (\pi^2/8G)\cdot H_0^2 \cdot R_0^3. \eqno(3)$$
For estimating $R_0$, we calculated for any galaxy with the distance
$D$ and radial velocity $V$ its spatial separation from NGC~253
$$  R^2 = D^2 + D^2_{N253} - 2D\cdot D_{N253}\cdot \cos\theta $$
and its projected radial velocity with respect to NGC~253

$$(V - V_{N253})_p = V\cdot\cos\lambda - V_{N253}\cdot\cos(\theta+\lambda),$$
 where $\theta$ is an angular distance of the galaxy from NGC 253, and
tan\,$\lambda = D_{N253}\cdot\sin\theta/(D - D_{N253}\cdot\cos\theta).$
Here we assumed that the peculiar velocities of the galaxies are small in
comparison with velocities of the regular Hubble flow. The estimated values
$(V - V_{N253})_p$ and $R$ for 15 galaxies around NGC~253 are presented
in column (9) of Table 2. The distribution of relative radial velocities
and spatial separations is shown in Fig. 6. Here galaxies with accurate
(``Cep'', ``RGB'', ``SBF'') and with rough (``TF'') distance estimates are indicated
by filled circles and crosses, respectively. As seen from Fig.\ 6, among
sufficiently distant galaxies with $R > 0.7 Mpc$ there are no galaxies
approaching NGC~253 ( the region of cosmological expansion). In particular,
NGC~7793 together with its companions move away from NGC~253 too. But
within $R < 0.7 Mpc$ there are galaxies both with negative (DDO~6) as well as
positive (NGC~247) velocities with respect to NGC~253 ("virialized" zone).
Based on these (still incomplete) data we can
conclude that the radius of the zero-velocity surface for
the NGC 253 group is  $R_0$= 0.7$\pm$0.1 Mpc. According to equation (3)
this corresponds to a total mass of $M_0$= (0.55$\pm$0.22) $10^{12} M_{\sun}$
or $M_0/L_B = (29\pm11) M_{\sun}/L_{\sun}$, which is 3--5 times lower than the
virial/orbital mass estimates derived above. Measurements of radial
velocities for the remaining four dSph galaxies, situated in the range
$R$= 0.47 -- 2.07 Mpc, will allow one to derive the radius $R_0$ and the
total mass of the NGC~253 group with higher accuracy.

\section {Concluding remarks}

Our measurements of accurate distances to 9 galaxies in Sculptor 
clarified the structure and kinematics of this nearby complex
of galaxies. However, nine other presumably nearby galaxies in the same area
(with $V_{LG} <$ 500 km s$^{-1}$ or $D < $5 Mpc) still remain without reliable
distance estimates ( NGC~55, NGC~625, NGC~247, SDIG, and ESO~149-03)
or radial velocity estimates (ESO~410--05, KDG~2, FG~24, and Sc~22).
Our new data on galaxy distances confirm the conclusion drawn by Jerjen
et al. (1998) that the so-called group in Sculptor is a loose
``cloud'' of galaxies of 1$\times$6 Mpc in size, extended along the line of sight.
Apparently, the near and the far parts of this ``cigar'' are not gravitationally
bound to each other, but take part in the general Hubble flow. In this sense, the
Sculptor cloud looks like another nearby cloud,  Canes Venatici I
(Karachentsev et al. 2003). Both these loose clouds are populated mostly
by dwarf galaxies, and their luminosity function has a flat ``primordial''
shape. As was noted by Jerjen et al.(1998) and Karachentsev et al.(2003), the
galaxy complexes in Sculptor and Canes Venatici together with the Local Group
form an amorphous filament extending over $\sim$10 Mpc.

  The nearby galaxy complex in Sculptor is a suitable case to study
the ``anatomy'' of virial mass excess. In projection onto the sky, the Sculptor
filament has a rather high overdensity, and can be easely identified
as a usual group by the ``friend of friends'' algorithm
(Huchra \& Geller 1982) or by the method of hierarchical trees (Materne 1978).
In his Nearby Galaxy Catalog, Tully (1988) denotes the Sculptor group by
the number ``14--13''. Based on radial velocities and projected separations
of 11 members of the group, Tully (1987) estimated its virial
mass-to-luminosity ratio to be $M_{vir}/L_B = 328 M_{\sun}/L_{\sun}$. There the
galaxies NGC~55, NGC~253, NGC~7793, and DDO~226 were considered as
members of a single group. Apart from real and probable members of the
Sculptor complex, Tully included in the group also the galaxy PGC~71145, whose
radial velocity is +16 km s$^{-1}$ (instead of +1600 km s$^{-1}$ as a result of
a misprint in Longmore et al. 1982). Comparing the ratio $M_{vir}/L_B$
from Tully (1987) with our estimate of the total mass-to-luminosity ratio,
29$\pm$11 $M_{\sun}/L_{\sun}$, we conclude that by using the more reliable
and precise
observational data that are now available, as well as our new approach to 
the determination of mass,
can decrease the total mass estimate of the group by one order of magnitude.

\acknowledgements
 The authors wish to thank C.Carignan, the referee, for his useful
comments.
{Support for this work was provided by NASA through grant GO--08601.01--~A
from the Space Telescope Science Institute, which is operated by the
Association of Universities for Research in Astronomy, Inc.,
under NASA contract NAS5--26555.
This work was partially supported by
RFBR grant 01--02--16001 and DFG-RFBR grant 02--02--04012.
D.G. gratefully acknowledges support from the Chile {\sl Centro de
Astrof\'\i sica} FONDAP No. 15010003.

 The Digitized Sky Surveys were produced at the Space Telescope
Science Institute under U.S. Government grant NAG W--2166. The
images of these surveys are based on photographic data obtained
using the Oschin Schmidt Telescope on the Palomar Mountain and the UK
Schmidt Telescope. The plates were processed into the present
compressed digital form with permission of these institutions.

 This project made use of the NASA/IPAC Extragalactic Database (NED),
which is operated by the Jet Propulsion Laboratory, Caltech, under
contract with the National Aeronautics and Space Administration.}

{}

%\end{document}

%\end{document}

\begin{table}
\caption{New distances to galaxies in Sculptor region. }
\begin{tabular}{|lcrrrrrrrr|} \hline

Name   &   RA (B1950) Dec  &  $a$  &  $B_T$  & $T$ & $V_h$  &$I(TRGB)$ & $(m-M)_0$& $D_{MW}$ &$(V-I)_{-3.5}$ \\
       &                   & $b/a$ &  $A_b$  &   & $V_{LG}$ &        &       &      &  [Fe/H]  \\
\hline
       &                   &     &      &   &     &        &       &      &          \\
Sc22   &   002121.0$-$245855 & 0.9 &17.73 & $-$3&     &  24.10 &  28.12&  4.21&   1.40   \\
       &                   & .78 & 0.06 &   &     &   0.21 &   0.23&  0.43&  $-$1.51   \\
       &                   &     &      &   &     &        &       &      &          \\
DDO226 &   004035.0$-$223127 & 2.2 &14.36 & 10&  357&  24.44 &  28.46&  4.92&   1.28   \\
       &                   & .36 & 0.07 &   &  408&   0.24 &   0.26&  0.58&  $-$1.96   \\
       &                   &     &      &   &     &        &       &      &          \\
N253   &   004506.9$-$253354 &26.7 & 7.92 &  5&  241&  23.97 &  27.98&  3.94&   1.54   \\
       &                   & .22 & 0.08 &   &  274&   0.19 &   0.21&  0.37&  $-$1.12   \\
       &                   &     &      &   &     &        &       &      &          \\
KDG2   &   004651.9$-$182048 & 1.2 &16.37 & $-$1&     &  23.65 &  27.66&  3.40&   1.37   \\
E540--030 &                 & .92 & 0.10 &   &     &   0.20 &   0.22&  0.34&  $-$1.61   \\
       &                   &     &      &   &     &        &       &      &          \\
DDO6   &   004721.0$-$211718 & 1.7 &15.19 & 10&  295&  23.60 &  27.62&  3.34&   1.25   \\
       &                   & .41 & 0.07 &   &  348&   0.13 &   0.16&  0.24&  $-$2.08   \\
       &                   &     &      &   &     &        &       &      &          \\
E540--032&  004756.0$-$201044 & 1.3 &16.44 & $-$3&     &  23.66 &  27.67&  3.42&   1.42   \\
FG24   &                   & .92 & 0.09 &   &     &   0.14 &   0.17&  0.27&  $-$1.45   \\
       &                   &     &      &   &     &        &       &      &          \\
E245--05&   014257.9$-$435054 & 3.8 &12.73 & 10&  394&  24.22 &  28.23&  4.43&   1.25   \\
P6430  &                   &0.89 & 0.07 &   &  308&   0.21 &   0.23&  0.45&  $-$2.08   \\
       &                   &     &      &   &     &        &       &      &          \\
UA442  &   234109.0$-$321412 & 6.4 &13.58 &  9&  267&  24.13 &  28.15&  4.27&   1.18   \\
       &                   & .14 & 0.07 &   &  299&   0.26 &   0.27&  0.52&  $-$2.40   \\
       &                   &     &      &   &     &        &       &      &          \\
N7793  &   235515.0$-$325206 & 9.3 & 9.70 &  7&  229&  23.95 &  27.96&  3.91&   1.50   \\
       &                   & .68 & 0.08 &   &  252&   0.22 &   0.24&  0.41&  $-$1.22   \\
       &                   &     &      &   &     &        &       &      &          \\
\hline
\end{tabular}
\end{table}
% \end{document}

\begin{table}
\caption{List of 21 nearby galaxies in the Sculptor group region.}
\begin{tabular}{|lcrrrrrrrl|} \hline

Name     & RA (B1950) Dec  &  $a$  &  $B_T$& $T$  &$V_h$  & $W_{50}^c$  & $D_{MW}$ & $dV_p$  & Notes          \\
	 &    $SGL \;\;\; SGB$   & $b/a$ &  $A_b$  &  & $V_{LG}$ & $M_b$   & $D_{LG}$ &  $R$   &                \\
\hline
	 &                 &     &      &  &     &       &      &      &                \\
E349--031 & 000540.9$-$345124 & 1.1 &15.48 &10& 207 &  35   & 4.1  &  44  & TF             \\
SDIG     &   260.18+0.40   & .82 & 0.05 &  & 216 &$-$12.65 & 4.02 & 0.90 & present paper  \\
	 &                 &     &      &  &     &       &      &      &                \\
N55      & 001238.0$-$392954 &32.4 & 8.84 & 9& 129 & 175   & 1.8  & 170  & TF             \\
	 &   256.25$-$2.36   & .17 & 0.06 &  & 111 &$-$17.50 & 1.78 & 2.26 & present paper  \\
	 &                 &     &      &  &     &       &      &      &                \\
N59      & 001253.0$-$214318 & 2.7 &13.12 &$-$3& 361 &       & 5.3  & 164  & SBF            \\
KK2      &   273.13+3.15   & .48 & 0.09 &  & 431 &$-$15.59 & 5.12 & 1.62 & Jerjen \&, 1998 \\
	 &                 &     &      &  &     &       &      &      &                \\
E410--005 & 001300.3$-$322728 & 1.3 &14.90 &$-$1&     &       & 1.92 &      & RGB            \\
KK3      &   262.95$-$0.26   & .77 & 0.06 &  &     &$-$11.58 & 1.85 &      & Kar. \&,2000    \\
	 &                 &     &      &  &     &       &      &      &                \\
Sc22     & 002121.0$-$245855 & 0.9 &17.73 &$-$3&     &       & 4.21 &      & RGB            \\
	 &   270.62+0.31   & .78 & 0.06 &  &     &$-$10.45 & 4.05 &      & present paper  \\
	 &                 &     &      &  &     &       &      &      &                \\
E294--010 & 002406.2$-$420756 & 1.1 &15.60 &$-$3& 117 &       & 1.92 & 195  & RGB            \\
	 &   254.37$-$5.27   & .64 & 0.02 &  &  81 &$-$10.84 & 1.85 & 2.06 & Kar. \&,2002  \\
	 &                 &     &      &  &     &       &      &      &                \\
DDO226   & 004035.0$-$223127 & 2.2 &14.36 &10& 357 &  51   & 4.92 & 134  & RGB            \\
	 &   274.23$-$3.21   & .36 & 0.07 &  & 408 &$-$14.17 & 4.74 & 1.01 & present paper  \\
	 &                 &     &      &  &     &       &      &      &                \\
N247     & 004439.6$-$210158 &21.4 & 9.86 & 7& 160 & 222   & 4.09 &  19  & TF             \\
	 &   275.92$-$3.73   & .32 & 0.08 &  & 215 &$-$18.28 & 3.90 & 0.36 & present paper  \\
	 &                 &     &      &  &     &       &      &      &                \\
N253     & 004506.9$-$253354 &26.7 & 7.92 & 5& 241 & 420   & 3.94 &   0  & RGB            \\
	 &   271.57$-$5.01   & .22 & 0.08 &  & 274 &$-$20.14 & 3.79 &   0  & present paper  \\
	 &                 &     &      &  &     &       &      &      &                \\
KDG2     & 004651.9$-$182048 & 1.2 &16.37 &$-$1&     &       & 3.40 &      & RGB            \\
E540--030 &   278.65$-$3.52   & .92 & 0.10 &  &     &$-$11.39 & 3.20 &      & present paper  \\
	 &                 &     &      &  &     &       &      &      &                \\
DDO6     & 004721.0$-$211718 & 1.7 &15.19 &10& 295 &  24   & 3.34 & $-$53  & RGB            \\
	 &   275.84$-$4.40   & .41 & 0.07 &  & 348 &$-$12.50 & 3.16 & 0.55 & present paper  \\
	 &                 &     &      &  &     &       &      &      &                \\
E540--032 & 004756.0$-$201044 & 1.3 &16.44 &$-$3&     &       & 3.42 &      & RGB            \\
FG24     &   276.95$-$4.24   & .92 & 0.09 &  &     &$-$11.32 & 3.23 &      & present paper  \\
	 &                 &     &      &  &     &       &      &      &                \\
N300     & 005231.8$-$375712 &21.9 & 8.95 & 7& 144 & 212   & 2.15 & 163  & Cep            \\
	 &   259.81$-$9.50   & .71 & 0.06 &  & 114 &$-$17.77 & 2.11 & 1.78 & Freedman \&,1992 \\
\end{tabular}
\end{table}

\begin{table}
\begin{tabular}{|lcrrrrrrrl|} \hline

	 &                 &     &      &  &     &       &      &      &                \\
N625     & 013254.9$-$414130 & 6.4 &11.59 & 9& 405 &  94   & 2.7: &  23  & TF             \\
	 &   257.26$-$17.74  & .28 & 0.07 &  & 335 &$-$15.64 & 2.69 & 1.65:& present paper  \\
	 &                 &     &      &  &     &       &      &      &                \\
E245--05  & 014257.9$-$435054 & 3.8 &12.73 & 9& 394 &       & 4.43 & 116  & RGB            \\
	 &   255.14$-$19.74  & .89 & 0.07 &  & 308 &$-$15.57 & 4.30 & 1.67 & present paper  \\
	 &                 &     &      &  &     &       &      &      &                \\
I5152    & 215926.6$-$513214 & 5.2 &11.06 &10& 124 & 116   & 2.07 & 212  & RGB            \\
E237--27  &   234.23+11.53  & .62 & 0.11 &  &  75 &$-$15.63 & 2.18 & 2.63 & Kar. \&, 2002   \\
	 &                 &     &      &  &     &       &      &      &                \\
KK258    & 223756.3$-$310340 & 1.6 &17.36 &$-$3&     &       &      &      &                \\
	 &   255.48+18.58  & .50 & 0.06 &  &     &       &      &      &                \\
	 &                 &     &      &  &     &       &      &      &                \\
UA438    & 232347.3$-$323957 & 1.5 &13.86 &10&  62 &  57   & 2.23 & 179  & RGB            \\
	 &   258.88+9.28   & .80 & 0.06 &  &  99 &$-$12.94 & 2.16 & 1.86 & Kar. \&,2002  \\
	 &                 &     &      &  &     &       &      &      &                \\
UA442    & 234109.0$-$321412 & 6.4 &13.58 & 9& 267 &  94   & 4.27 &  82  & RGB            \\
	 &   260.78+6.12   & .14 & 0.07 &  & 299 &$-$14.64 & 4.01 & 1.15 & present paper  \\
	 &                 &     &      &  &     &       &      &      &                \\
E149--003 & 234925.5$-$525121 & 2.2 &15.0  &10& 577 &  56   & 6.4  & 293  & TF             \\
	 &   242.31$-$3.23   & .18 & 0.06 &  & 501 &$-$14.10 & 6.44 & 3.54 & present paper  \\
	 &                 &     &      &  &     &       &      &      &                \\
N7793    & 235515.0$-$325206 & 9.3 & 9.70 & 7& 229 & 237   & 3.91 &  60  & RGB            \\
	 &   261.30+3.12   & .68 & 0.08 &  & 252 &$-$18.34 & 3.82 & 0.90 & present paper  \\
\hline
\end{tabular}
\end{table}
%\end{document}
\begin{table}
\caption{Properties of nearby groups in Sculptor.}
\begin{tabular}{|lcccccccc|} \hline
   Group     &  $N$ & $ D $& $< R_p>$ & $\sigma_V$ & $L_B$     &$M_{vir}/L_B$ &$M_{orb}/L_B$ &$T_{cross}$ \\
	     &    &  Mpc &   kpc  &  km s$^{-1}$ &   $10^{10}$  $L_{\sun}$ &$M_{\sun}/L_{\sun}$   &$M_{\sun}/L_{\sun}$   & Gyr    \\
	      \hline
	     &    &      &        &         &        &        &          &           \\
N300, N55,   &  4 & 1.95 &   279  &   15    & 0.33   &   54   &     45   &   18.6    \\
E294--10,     &    &      &        &         &        &        &          &           \\
E410--05      &    &      &        &         &        &        &          &           \\
	     &    &      &        &         &        &        &          &           \\
N253, N247,  &  6 & 3.94 &   370  &   54    & 1.91   &  143   &     83   &    6.9    \\
DDO6, Sc22,  &    &      &        &         &        &        &          &           \\
KDG2, FG24   &    &      &        &         &        &        &          &           \\
	     &    &      &        &         &        &        &          &           \\
N7793, UA442,&  3 & 3.91 &   205  &   34    & 0.30   &  260   &    140   &    6.0    \\
E349--31      &    &      &        &         &        &        &          &           \\
\hline
\end{tabular}
\end{table}
\newpage
\begin{figure*}
\centering
\vspace{5mm}
\includegraphics[width=12cm]{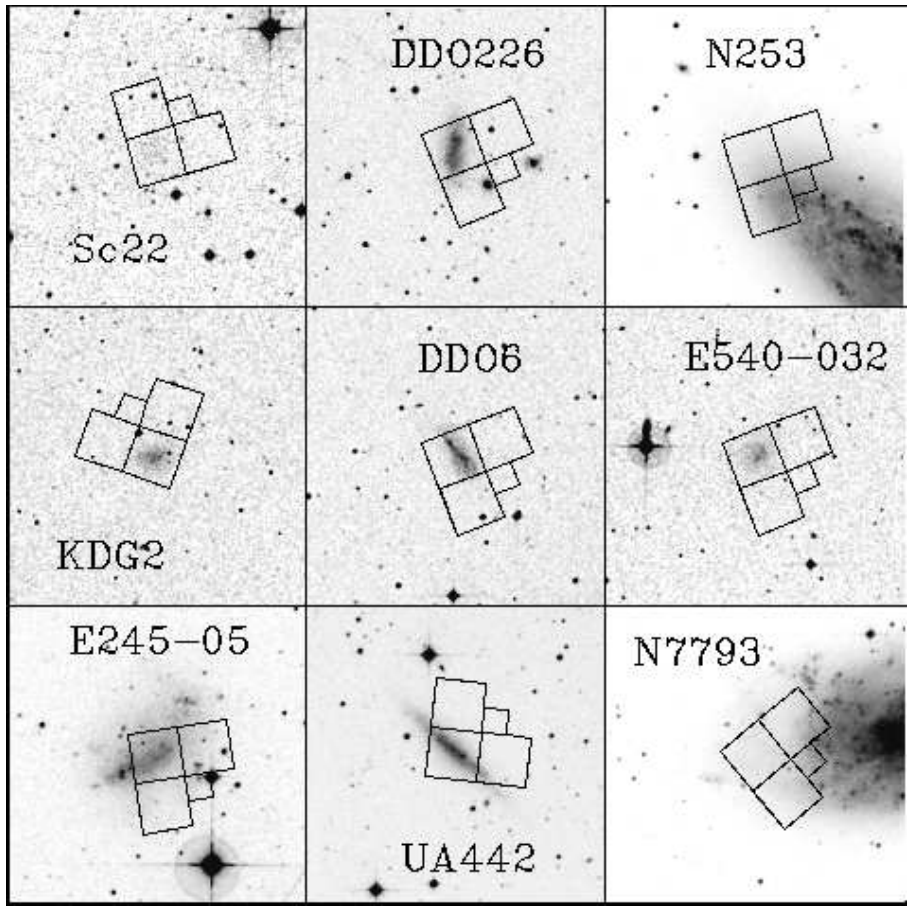}
\vspace{5mm}
\caption{Digital Sky Survey images of 9 nearby galaxies in Sculptor.
The field size is 8$\arcmin$, North is up and East is left.
The HST WFPC2 footprints are superimposed.}
\end{figure*}
\clearpage

\begin{figure*}
\caption{{\bf Top}: WFPC2 images of nine galaxies: Sc~22, DDO~226, NGC~253,
KDG~2, DDO~6, ESO~540--032, ESO~245--05, UGCA~442, and NGC~7793 produced by
combining the two 600s exposures obtained through the F606W and
F814W filters. The arrows point to the North and the East.
{\bf Bottom left}: The color-magnitude diagrams from the WFPC2 data for
the nine galaxies in Sculptor.
{\bf Bottom right}: The Gaussian-smoothed $I$-band luminosity function
restricted to red stars (top), and the
output of an edge-detection filter applied to the luminosity function
for the nine galaxies.}
\end{figure*}

\begin{figure*}
\centering
\vspace{5mm}
\includegraphics[width=15cm,angle=-90]{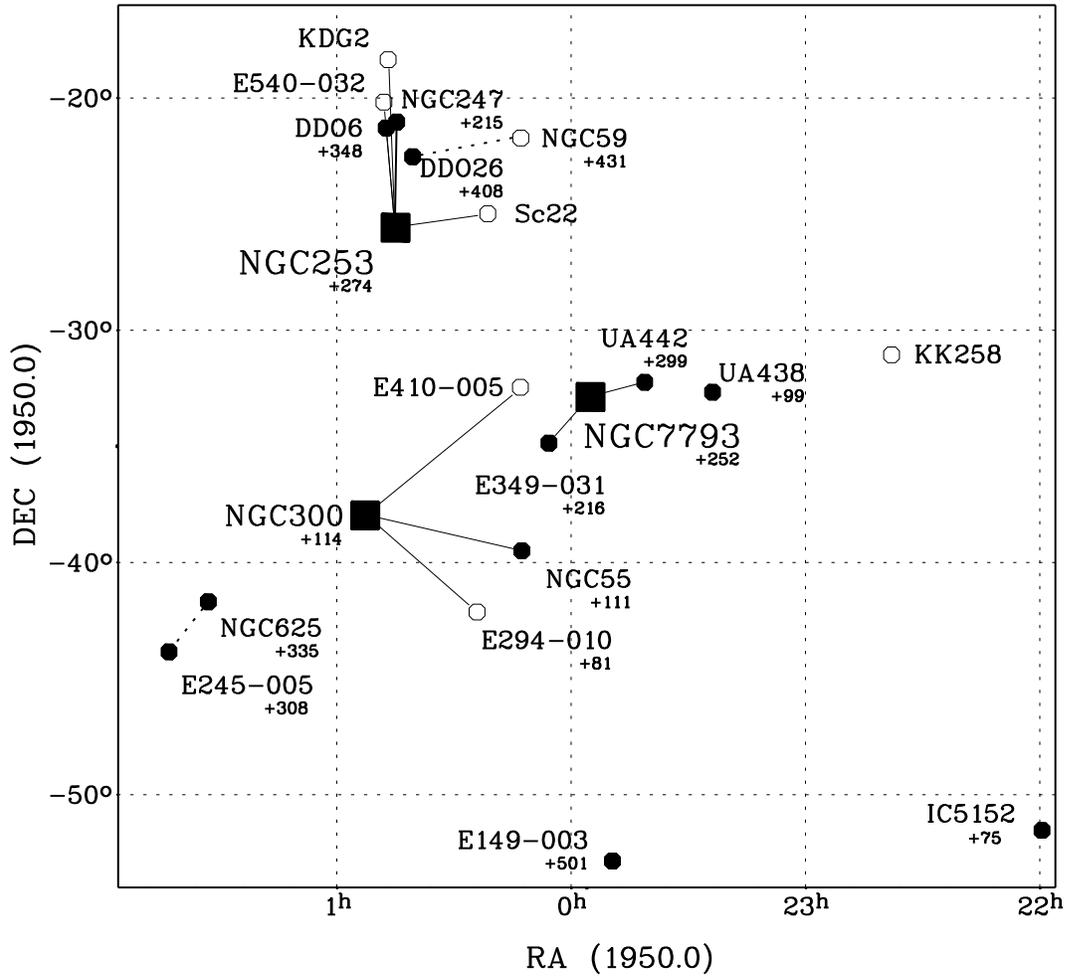}
\caption{Sky distribution of nearby galaxies in the direction of the
Sculptor group. Filled and open circles indicate dwarf irregular and
dwarf spheroidal galaxies, respectively. Large squares correspond to
the most luminous galaxy in each group. Their probable companions are
connected to the principal galaxies with straight lines. Radial velocities
of the galaxies relative to the Local Group centroid are indicated by
small numbers.}
\end{figure*}

\begin{figure*}
\centering
\vspace{5mm}
\includegraphics[width=12cm,angle=-90]{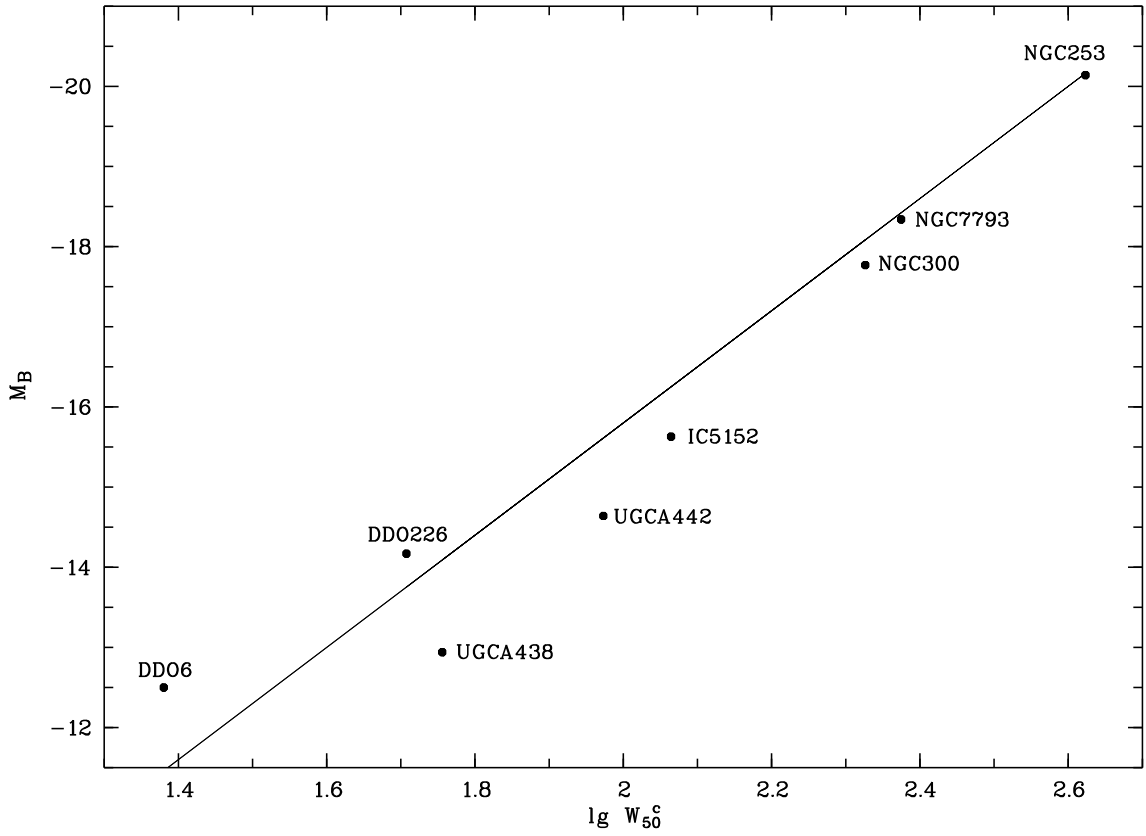}
\caption{Tully-Fisher relation for eight galaxies in Sculptor with
accurate distances.}
\end{figure*}
\clearpage
\begin{figure*}
\includegraphics[width=12cm,angle=-90]{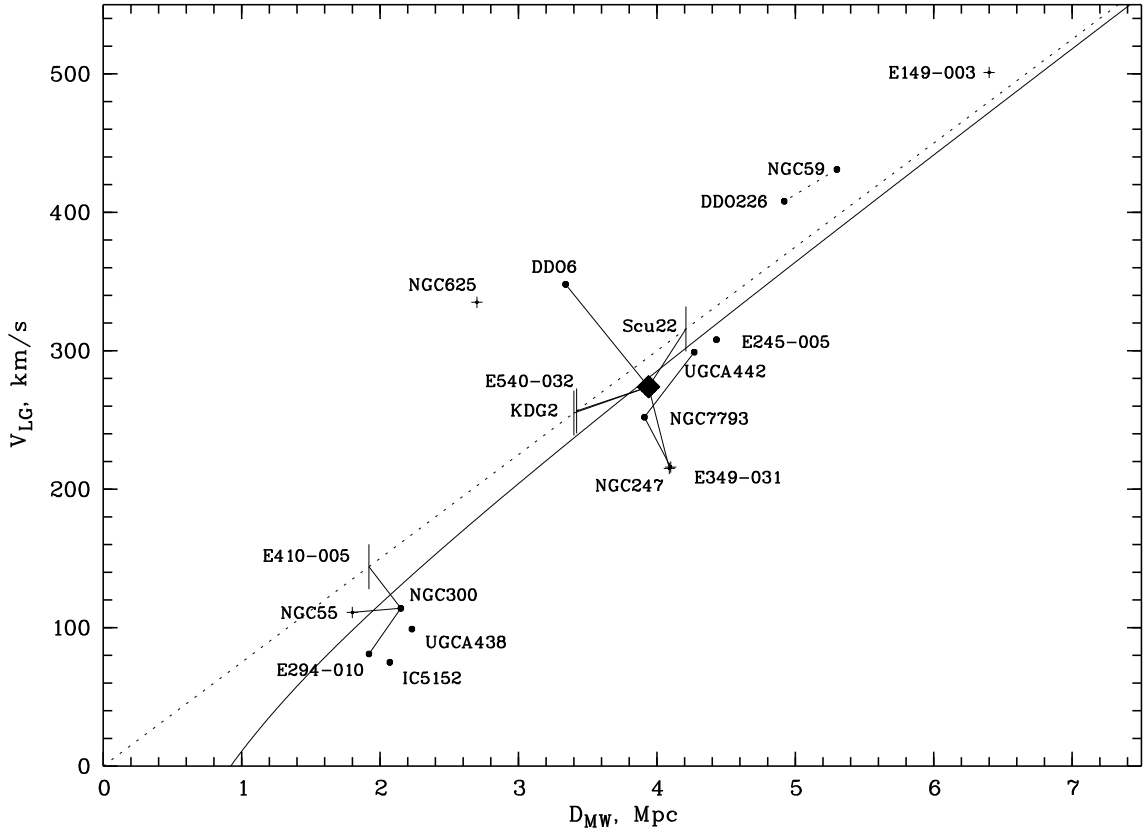}
\vspace{5mm}
\caption{Radial velocity -- distance relation for 20 nearby galaxies
in Sculptor. The galaxies with accurate distance estimates are shown
as filled circles, and galaxies with T-F distances are shown as
crosses. Four dSph galaxies without radial velocities are indicated
by vertical bars. The brightest galaxy, NGC~253, is indicated by a square.
The solid line corresponds to the Hubble relation with H = 75 km s$^{-1}$ Mpc$^{-1}$,
curved at small distances assuming a decelerating gravitational action
of the Local Group with a total mass of $1.3\cdot10^{12} M_{\sun}$.
Probable companions of the most luminous galaxies are connected with
these by straight lines.}
\end{figure*}
\begin{figure*}
\centering
\vspace{5mm}
\includegraphics[width=12cm,angle=-90]{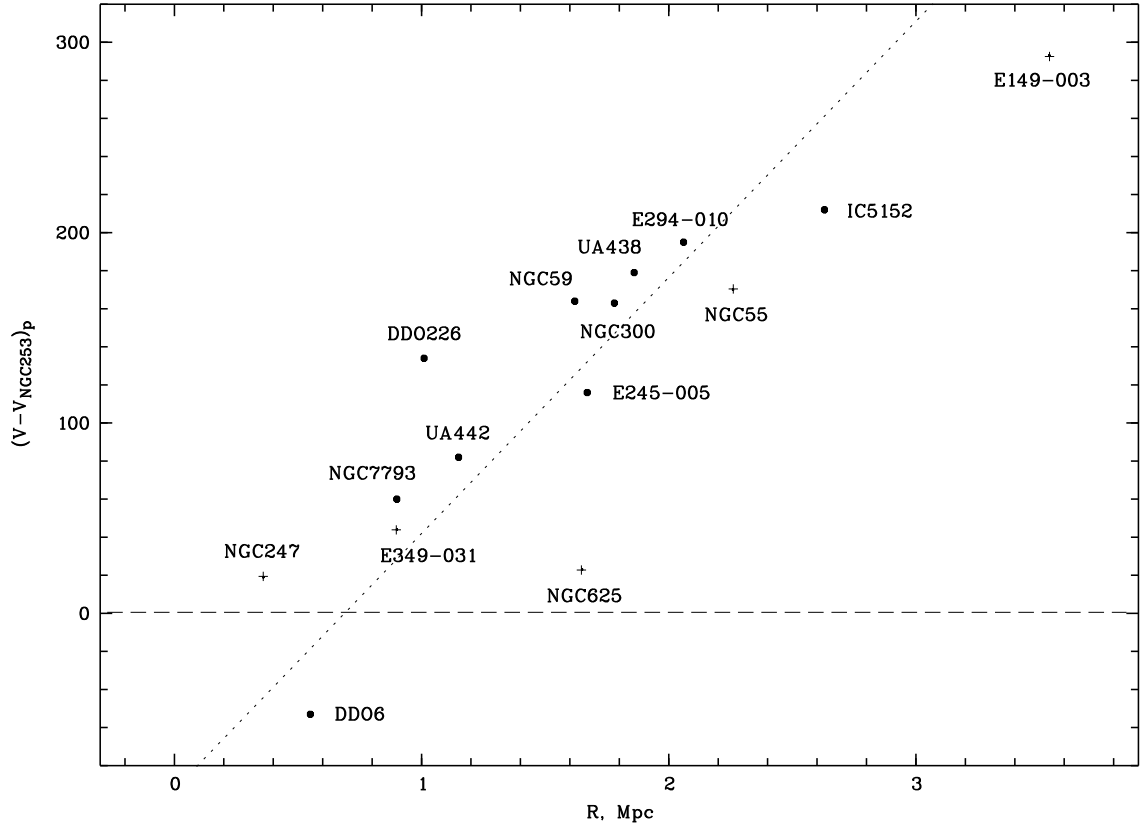}
\vspace{5mm}
\caption{ The distribution of the radial velocity difference and the
distance of nearby galaxies with respect to NGC~253. These data
yield a radius of the zero-velocity surface of $R_0$= 0.7 Mpc. }
\end{figure*}
%\end{document}
\end{document}